\documentclass[aip,jap,numerical,amsmath,amssymb]{revtex4-1}

\usepackage{graphicx}
\usepackage{dcolumn}% Align table columns on decimal point
\usepackage[export]{adjustbox}
\graphicspath{ {Figures/} }
\usepackage{fnpct}
\usepackage[colorlinks]{hyperref}
\usepackage{amsmath}
\usepackage{amsxtra}
\usepackage{amstext}
\usepackage{amssymb}
\usepackage{latexsym}
\usepackage{dsfont}
\usepackage{textcomp}
\usepackage{ulem}
\usepackage{siunitx}
\usepackage[capitalise]{cleveref}
\usepackage{multirow}
\usepackage{color}
\usepackage{epstopdf}
\usepackage{caption, subcaption}
\usepackage{bm}
\usepackage{soul}
\usepackage{natbib}
\usepackage{hyperref}
\usepackage[english]{babel}
\usepackage{lipsum}
\usepackage{commath}
\usepackage{graphicx}% Include figure files
\usepackage{dcolumn}% Align table columns on decimal point
\usepackage{bm}% bold math
\usepackage{mathtools,upgreek}
\usepackage{tabularx,ragged2e,booktabs,caption}
\usepackage{makecell}
\usepackage{bigdelim,bigstrut}
\newcolumntype{C}[1]{>{\Centering}m{#1}}

\renewcommand{\vec}[1]{\bm{\mathrm{#1}}}
\newcommand{\etal}{\textit{et~al}. }
\hypersetup{
	colorlinks=true,
	linkcolor=blue,
	filecolor=blue,      
	urlcolor=blue,
}

\begin{document}
\newcommand{\rood}[1]{\textcolor{red}{[#1]}} %for displaying red texts
\newcommand{\gre}[1]{\textcolor[rgb]{0,0.35,0}{[#1]}} %for displaying green texts
\newcommand{\blu}[1]{\textcolor{blue}{#1}} %for displaying red texts
\newcommand{\pdx}[1][]{\pd{x}#1}
\newcommand{\pdt}[1][]{\pd{t}#1}
\newcommand{\eps}{\varepsilon}
\newcommand{\sig}{\sigma}
\newcommand{\kap}{\kappa}
\newcommand{\gam}{\gamma}
\newcommand{\om}{\omega}
\newcommand{\bbet}{\bm{\beta}}
\newcommand{\bpsi}{\bm{\psi}}
\newcommand{\bkap}{\bm{\kap}}
\newcommand{\Vol}{\mathcal{V}}
\newcommand{\Sur}{\mathcal{S}}
\newcommand{\TM}{\mathsf{T}}
\newcommand{\SM}{\mathsf{S}}
\newcommand{\Oh}{\mathcal{O}}
\newcommand{\pt}{\mathcal{PT}}
\newcommand{\im}{\mathrm{i}}

\title{Angle-dependent Phononic Dynamics for Deep Learning and Source Localization}
\author{Weidi Wang}
\affiliation{Department of Mechanical Engineering, University of Massachusetts, Lowell, 
	Lowell, MA 01854, USA
}%
\author{Amir Ashkan Mokhtari}
\author{Ankit Srivastava}
\affiliation{Department of Mechanical, Materials, and Aerospace Engineering, Illinois Institute of Technology, Chicago, IL 60616, USA
}%
\author{Alireza V. Amirkhizi$^{1,}$}
\email{alireza\_amirkhizi@uml.edu}

\date{\today}
\begin{abstract}
In this work, a parameterized eigenvalue problem is analyzed for a phononic array in a 2D stress wave scattering setup, and a corresponding sensing application of this system is proposed to achieve source angle localization. The phononic domain consists of a periodic micro-structured medium, of which the eigen-wavevector band structure and the eigen-modes are exploited. The eigen-modes are naturally angle dependent due to changes in phases and periodic mode shapes determined by the incident angle. An intriguing aspect found in the band structure is that it exhibits angle-dependent transitions at the exceptional points (EPs) and critical angles (CAs), where the eigenvalues coincide or vanish. Coupled with these transitions, it is found that the eigen-modes switch their energy characteristics and symmetry patterns at these branch points, leading to enhanced angle dependence. Moreover, these eigen-modes also serve as the basis functions of the scattered waves. Therefore, the scattering response of the medium inherently possesses the angle-dependent properties, making this system naturally suitable for sensing applications. An artificial neural network (ANN) is trained with randomly weighted eigen-modes to achieve deep learning of the eigen features and angle dependence. The training data is derived only based on the eigen-modes of the unit cells. Nevertheless, the trained ANN can accurately identify the incident angle of an unknown scattering signal, with minimal side lobe levels and suppressed main lobe width. The ANN shows superior performance in comparison with standard delay-and-sum technique of estimating angle of arrival. The proposed application of ANN and micro-structured media highlights the physical importance of band structure topology and eigen-modes to a technological application, adds extra strength to the existing localization methods, and can be easily enhanced with the fast-growing data-driven techniques. 
\end{abstract}

\maketitle

	{\bf HIGHLIGHTS} 

\begin{itemize}
    \item Elucidation of the eigen-mode properties (particularly the symmetry patterns) in a phononic problem which exhibit EPs and CAs;
    \item Development of a deep ANN which is trained by the phononic eigen-modes and can provide near perfect source localization results.
\end{itemize}

\section{Introduction}

Traditional angle-of-arrival (AoA) measurements rely on time-of-flight calculation or finding maximum signal strength during antenna rotation\cite{Chiariotti2019}, and their precision is limited especially when strong noise is present. Among the traditional phased array approaches, it has been shown that sensor arrays made of non-homogeneous material will provide extra information for localization calculations and effectively enhance the sensing directivity\cite{Joffre2020}. Non-homogeneous media, especially mechanical metamaterials (MMs) and phononic crystals (PCs), exhibit exotic properties associated with wave propagation due to the collective or local behavior of their micro-structures. These micro-structured media can lead to unique features such as wave attenuation\cite{Nemat-Nasser2015,Aghighi2019a,DAlessandro2019a,Elmadih2021}, topological insulation\cite{Guo2017} and angle-dependent dynamic properties\cite{Amirkhizi2018c}.

\begin{figure}[!ht]
		\centering\includegraphics[height=110pt]{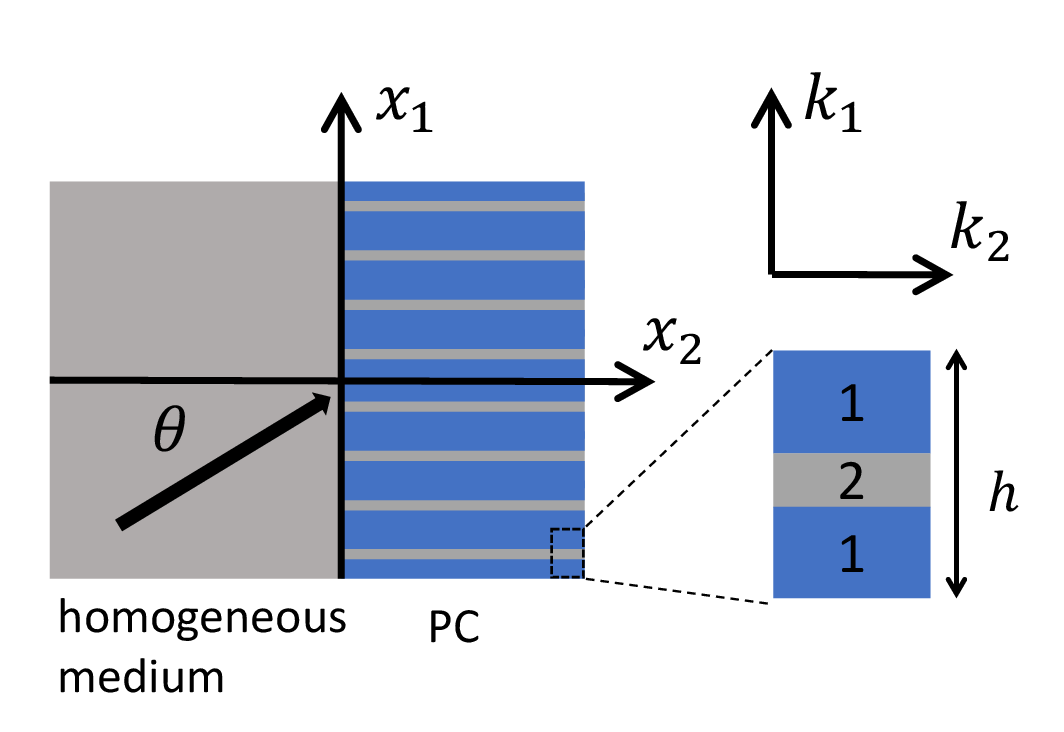}
		\caption{The setup of 2D stress wave scattering at the interface between homogeneous medium and an array of phononic crystals. Left and right domains are semi-infinite. The incident oblique wave is denoted by the arrow with an angle~$\theta$ measured from the normal to the interface.\label{fig:setup}}
\end{figure}

In this work, we introduce the use of eigen-modes of a micro-structured medium and exploit its angle-sensitive nature applicable for the purpose of AoA estimation. 
As the most fundamental aspect for any AoA estimation algorithm, the scattering response of any system (homogeneous or heterogeneous) is dependent on the wavevector. However, in specifically designed periodic media, the presence of branch points leads to stronger qualitative dependence on the angle, most easily demonstrated in the modal symmetry attributes. We will seek to take advantage of this attribute in this paper to propose a more robust source localization methodology. To show this, a periodic array of phononic crystals (PC) is used as the sensor medium to collect signals of 2D oblique stress waves incident from a homogeneous medium, see~\cref{fig:setup}. The wavevector~$\vec{k}$ in this oblique scattering problem includes two components:~$k_1$ parallel to the interface and~$k_2$ normal to the interface. The band structure and eigen-modes of the PC are first analyzed.

The band structure of a MM or PC is usually represented by~$\omega(\vec{k})$, i.e., eigenfrequencies as functions of wavevector, such as in Ref.\cite{Amirkhizi2018d}. The~$\omega(\vec{k})$ band structure provides information on wave speeds and frequency stop bands in a desired direction. For linear elastic materials and real-valued wavevector, the equation of motion is self-adjoint (Hermitian) and the solved eigenfrequency will be real-valued. An alternative form of band structure may be arrived at by solving for the eigen-wavevector component~$k_2(\omega,k_1)$, and is useful for finding the oblique scattering wave field when a wave is incident at the interface between two domains\cite{Srivastava2017a}. In such problems, the frequency~$\omega$ and wavevector component~$k_1$ are prescribed as real values based on the incident wave, and one solves for the complex~$k_2$ wavevector normal to the interface. A detailed analysis on the~$k_2(\omega,k_1)$ eigenvalue problem is formulated by Mokhtari \etal\cite{Mokhtari2019a}. In the example in this work, the wavevector band structure~$k_2(\omega,k_1)$ of the phononic crystal is pre-calculated in the~$k_2(\omega,k_1)$ eigenvalue analysis, in which the parameter~$k_1$ is related to the incident angles through Snell’s law in the incident (homogeneous) domain. This representation provides the critical opportunity to express the reflected/transmitted scattering signal at the interface as a weighted sum of the eigen-modes. %Therefore, not only do the modal features affect the measured scattering signal, the particular partition of the scattered wave into these modes is also a significant function of the incident wavevector.

Two types of branch points are identified in this representation of PC band structure. Both types of branch points have angle-sensitive natures and are the spectral boundaries between propagating modes and evanescent modes. The first type of branch points is related to the critical angles\cite{Rokhlin1989} (CAs), where one wavevector solution transitions from purely real to imaginary (or vice versa). The modes associated with the CAs have zero wavevector component normal to the boundary, which leads to total internal reflection\cite{Abraham2019}. 
%\rood{need some refs}
% CA literature
The second type of branch points is referred to as exceptional points (EPs)\cite{Heiss}, which are spectral singularities in a parametric non-Hermitian systems. At the EPs, two modes coalesce with identical eigenvalues and eigen-modes.
Mode coalescence often appears in literature mostly as a mathematical and abstract concept because accessing these EPs are physically difficult (e.g. requires gain units in PT symmetric medium\cite{Zhu2014}). Any infinitesimally small parametric perturbation at an EP will separate the coalesced modes into independent ones and induces an abrupt phase transition. In an elasto-dynamic setup\cite{Lustig2019b} similar to this work, unusual energy transport is found at the EPs.
% EP literature
It has been shown that the unique topology of the bands structure near the EP degeneracy leads to enhanced sensitivity\cite{Shmuel2020}, therefore, it may be used to develop sensing devices in various physical setups\cite{Wiersig2014a,Wiersig2016,Djorwe2019a}.

In the present study, spontaneous symmetry breaking is found to occur at the CAs and the EPs, where the modes transition between propagating bulk modes which exhibit symmetric eigenfunctions and zero-energy edge modes which have asymmetric eigenfunctions. Since the eigen-modes may be used as basis functions of the scattered wave, the analysis can be effectively conducted within the subspace spanned by the dominant modes. Such a subspace inherently has angle-dependent features of the eigen-modes. This motivates the use of the spanned subspace as a means to improve estimation of angle of arrival. The unique band topology and exotic modal behavior of the micro-structured medium deserve investigations not only with a physics-based perspective but also a data-driven/machine learning approach: the former explains the physical nature of the problem while the latter learns and utilizes the underlying patterns for practical applications. 
Machine learning (ML) methods have shown strong capability for multiple applications such as regression and classification in a variety of fields. Niu~\etal\cite{Niu2017} showed the promising potential of using three machine learning methods (neural network (NN), support vector machine (SVM), and random forest (RF)) for estimating acoustic source ranges.
Artificial neural networks (ANNs), particularly among the ML methods, have the potential to learn the hidden mechanisms and approximate complicated input-output relations. The related application scenarios of deep learning (DL) networks include layout design\cite{Zhang2019a} and band structure computation\cite{Finol2019,Liu2019}.
In the present study, a deep NN is constructed to be capable of connecting the modal features with the incident angle, thus providing an effective tool for estimation of the bearing angle. Hence, we demonstrate the sensing potential based on deep learning of the angle-dependent properties of the phononic modes. The source localization application is approached as a supervised multi-label classification problem. The ANN is constructed and trained as the source angle classifier. The eigen-modes associated with each angle are randomly weighted and summed to serve as the training input. With a large enough set of these training samples, the ANN is able to learn and identify the angle-dependent features of the subspace spanned by the eigen-modes. Although the scattering signal is unknown to the trained NN, the abstract features of its underlying subspace associated with each angle have been fed into the NN. Therefore, the ANN can accurately identify the incident angle of an unknown scattering signal. This approach is shown to have major improvements compared with conventional localization algorithms such as delay-and-sum, in terms of the main lobe width and the side lobe levels.

%The pre-calculated mode shape information also have the potential to be utilized for finding optimal sensor locations through data-driven techniques\cite{Manohar2018} such as compressed sensing\cite{Brunton2015} and principal component analysis\cite{Brunton2016a}.

This paper is organized as follows. First, we briefly revisit the~$k_2(k_1,\omega)$ eigenvalue formulation for the oblique scattering problem. Then, we discuss the band structure topology and the eigen-mode properties of the PC array, with particular focus on the symmetry breaking at the branch points. 
Then, the scattering signal formulation is presented. 
Finally, we introduce the proposed localization method using ANN. The preparation process of the training, validation, and signal data sets will be stated. The ANN architecture and the training process will be described. Finally we discuss the sensing performance of the trained NN and compare it with a traditional localization algorithm.

\section{Band structure and eigen-mode analysis}\label{sec:band}

The studied scattering problem is formulated similar to our previous work\cite{Mokhtari2020a} except for the different frequency range and unit cell size used here. An in-plane stress wave is incident from a semi-infinite domain of homogeneous medium to an array of PCs with an incident angle~$\theta$ normal to the interface, as illustrated in \cref{fig:setup}.

For the plane wave propagating problem in the two semi-infinite domains, it is advantageous to know a priori the band structure that characterizes the dynamics of each domain. The displacement and stress solutions in one unit cell have the form
\begin{equation}\label{eq:dispsolution}
    u_{i} (x_1)=\bar{u}_{i}(x_1)\exp[\im (k_1 x_1+k_2 x_2-\omega t)]
\end{equation} and
\begin{equation}\label{eq:stresssolution}
    \sigma_{ij} (x_1)=\bar{\sigma}_{ij}(x_1)\exp[\im (k_1 x_1+k_2x_2-\omega t)].
\end{equation}
Here~$u_{i}$ is the displacement components,~$\sigma_{ij}$ is the~$ij$ component of stress tensor,~$k_2$ is the wavevector component normal to the interface, and~$\omega$ is the angular frequency. The barred quantities are the periodic parts within one unit cell. The wavevector component~$k_1$ is parallel to the interface and is hence related to the incident angle~$\theta$ through Snell's law
\begin{equation}
    k_{in}\sin\theta=k_1
\end{equation}
where~$k_{in}=\omega/C$ is the wavevector of the incident wave, and~$C$ is the incident wave speed in the homogeneous medium (longitudinal or shear). The continuity at the interface between two domains requires that for each angle of incidence, a real-valued~$k_1 h \mod 2\pi$ is prescribed in the non-Hermitian eigenvalue problem, from which an infinite number of~$k_2$ eigenvalues can be found. With~$\vec{\gamma}=[k_1,0]$ and~$\vec{n}=[0,k_2]$, the the~$k_2(\omega,k_1)$ eigenvalue problem is formulated as:
\begin{equation}\label{eq:eig}
    \vec{A}\vec{\bar{\phi}}=k_2\vec{B}\vec{\bar{\phi}},
\end{equation}
where the mode shape is described by~$\vec{\bar{\phi}}=[\vec{\bar{u}},\vec{\bar{\sigma}}]^\top$, and
\begin{align}
\begin{split}
    \vec{A}&=\begin{pmatrix}
    \omega^2 \rho() & \nabla\cdot()-\im ()\cdot\vec{\gamma}\\
    -\vec{C}:\nabla()+\im\vec{C}:()\otimes \vec{\gamma} &\vec{I}
    \end{pmatrix},\\
    \vec{B}&=\begin{pmatrix}
    \vec{0}& \im()\cdot \vec{n}\\
    -\im \vec{C}:()\otimes\vec{n}&\vec{0}
    \end{pmatrix}.
    \end{split}
\end{align}
The bold symbols~$\{\vec{\bar{u}},\vec{\bar{\sigma}}\}$ represent quantities as vectors which include the in-plane components, $\vec{C}$ is the elasticity modulus, and $\rho$ is the density.
The details of this eigenvalue problem can be found in Ref\cite{Mokhtari2019a}. In the solutions considered here, the frequency~$f=\omega/(2\pi)=\SI{1.8}{kHz}$ is constant and the wavevector component~$k_2$ values are solved as complex eigenvalues for the varying parameter~$k_1$.

\subsection{Band structure}
\begin{figure}[h!]
	\begin{subfigure}[b]{0.5\linewidth}
		\centering\includegraphics[height=170pt]{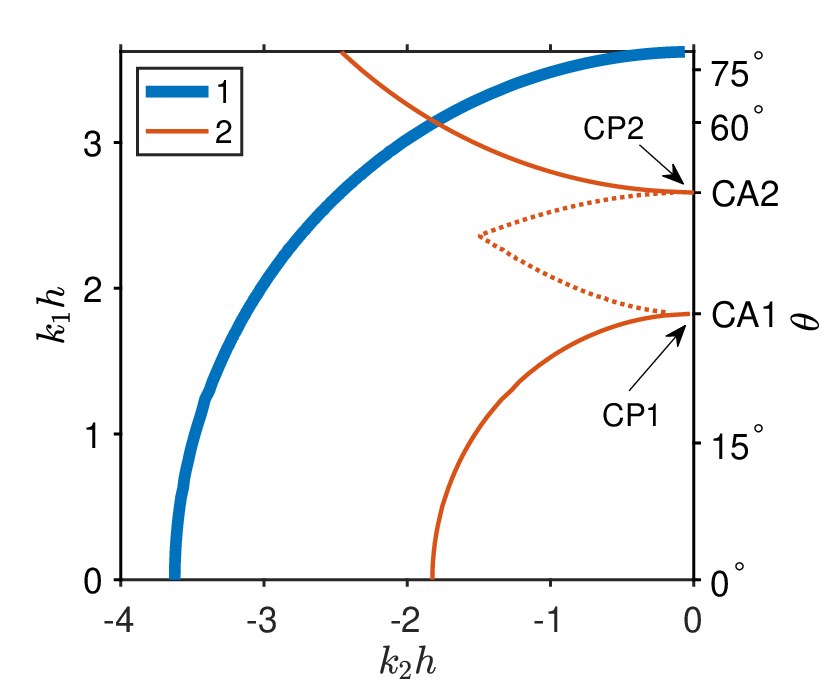}
		\caption{homogeneous\label{fig:refband}}
	\end{subfigure}%
	\begin{subfigure}[b]{0.5\linewidth}
		\centering\includegraphics[height=170pt]{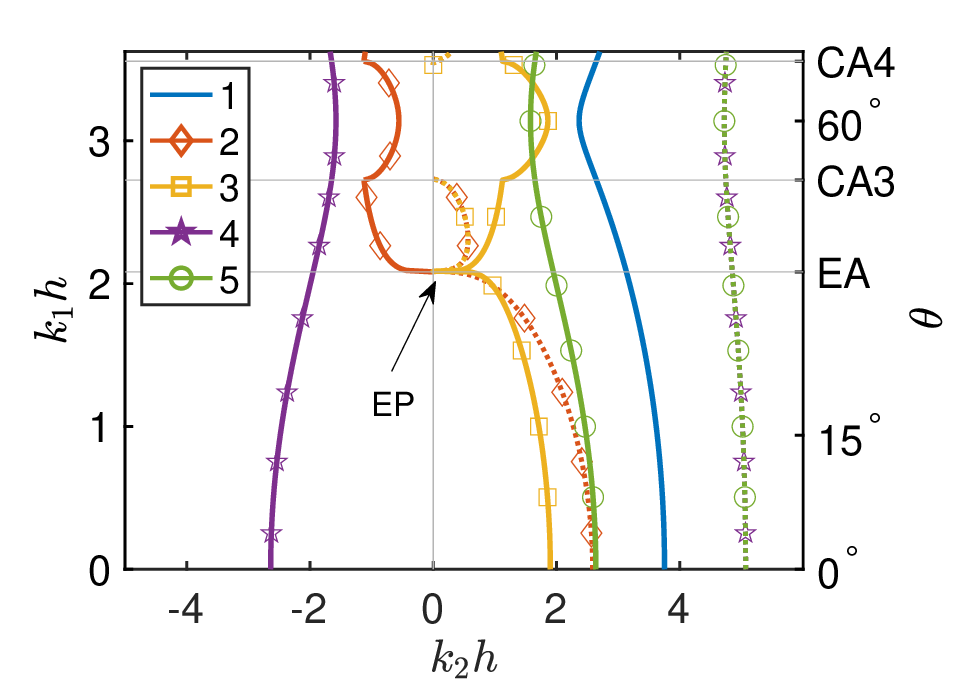}
		\caption{PC\label{fig:transband}}
	\end{subfigure}
	\caption{Wavevector band structures of the reflected waves in the homogeneous domain (\subref{fig:refband}) and the transmitted waves in the PC (\subref{fig:transband}). The~$k_2h$ values have real (solid) and imaginary (dotted) components. Only the first several important modes are shown. (colored online)\label{fig:band} }
\end{figure}

The band structures of the homogeneous domain and the PC domain are partially shown in \crefrange{fig:refband}{fig:transband}, respectively. The complete spectra will have symmetry with respect to~$k_2=0$. Only the modes that have physical meanings in the scattering problem are considered and shown here. 
A physically feasible eigen-mode of the homogeneous medium, representing the reflected wave solution, must (1) have non-positive energy flux vector component $F_2$ and (2) have~$\Im k_2\leq0$. Similarly, to represent a transmitted wave solution, a PC eigen-mode must (1) have non-negative energy flux vector component $F_2$ and (2) have~$\Im k_2\geq0$. The modes violating these requirements are not taken into consideration. The (1) requirements ensures that the reflected/transmitted waves transfer energy away from the interface. The (2) requirements prevent infinitely large amplitudes at~$x_2=\pm\infty$, given the solution forms~\crefrange{eq:dispsolution}{eq:stresssolution} .
%\rood{Can we provide reasoning similar to what I mentioned for homogeneous domain as well?} 
For each domain, only several modes with lowest~$|\Im k_2|$ values are shown in the band structure. Each of the solved eigen-modes represents a wave whose spatial features in~$x_2$ are determined by its~$k_2$ wavevector. A mode that has a real~$k_2$ eigenvalue will propagate and carry energy in~$x_2$ direction. A mode with complex eigenvalue, on the other hand, only allows the wave to propagate along the interface and is evanescent in~$x_2$ direction. As the incident angle varies, certain modes can transition from propagating to evanescent and vice versa.

In the reflected solutions \cref{fig:refband}, mode 1 represents the shear vertical wave reflected from the interface, and is a propagating mode for any incident angle~$\theta$. The reflected mode 2 contains three branches and two critical angles (CA1 and CA2) in between them. The real-valued branch from~$0^\circ$ to~$30.3^\circ$ is the longitudinal wave that can propagate in the bulk of the homogeneous medium. The purely imaginary branch from~$30.3^\circ$ to~$47.2^\circ$ is the surface mode that does not allow energy flux along~$x_2$. The third branch, corresponding to reflection angles from~$47.2^\circ$ to~$89^\circ$, is the shear mode of the second Brillouin zone, and has real~$k_2$ eigenvalues. The two critical angles (CAs) at~$30.3^\circ$ and~$47.2^\circ$ are boundaries between propagating and evanescent modes.

The band structure of transmitted waves in~ \cref{fig:transband} has more spectral features due to the micro-structure of the layered medium. Mode 1 has purely real eigenvalues and is thus a propagating mode. Mode 2 and 3 coalesce at the exceptional point (labelled as EP), associated with the exceptional angle (EA: 35.1$^\circ$). Two critical angles can be found in mode 2 and 3 (CA3: 48.8$^\circ$, and CA4: 78.4$^\circ$). 
These special angles are associated with emergence or annihilation of energy-carrying branches. At the EP, the eigenvalues and eigenfunctions of the two modes (2 and 3) will be identical. From~$0^\circ$ incidence to~$35.1^\circ$ incidence, mode 2 has purely imaginary eigenvalues while mode 3 has real eigenvalues. At the EP, the eigenvalues of the two modes are identical and close to zero. From EA (35.1$^\circ$) to CA3 (48.8$^\circ$), the eigenvalues of mode 2 and 3 share the same imaginary parts while their real parts are negatives of each other. Mode 2 and 3 become propagating from CA3 to CA4 (78.4$^\circ$) and have distinct real-valued eigenvalues. For angles from CA4 to 89$^\circ$, mode 2 and 3 again possess complex-valued eigenvalues. For the entire angle range, mode 4 and 5 share the same~$\Im k_2 h$ (the dotted lines overlapped as the right-most curve in \cref{fig:transband}). Their real parts of~$k_2h$ are negatives of each other. Due to their complex-valued~$k_2$ eigenvalues, mode 4 and 5 are evanescent modes. 

\subsection{Modal symmetry and energy flux}

\begin{figure}%[!ht]
\centering
	\begin{subfigure}[b]{1\linewidth}
		\centering\includegraphics[width=0.92\linewidth]{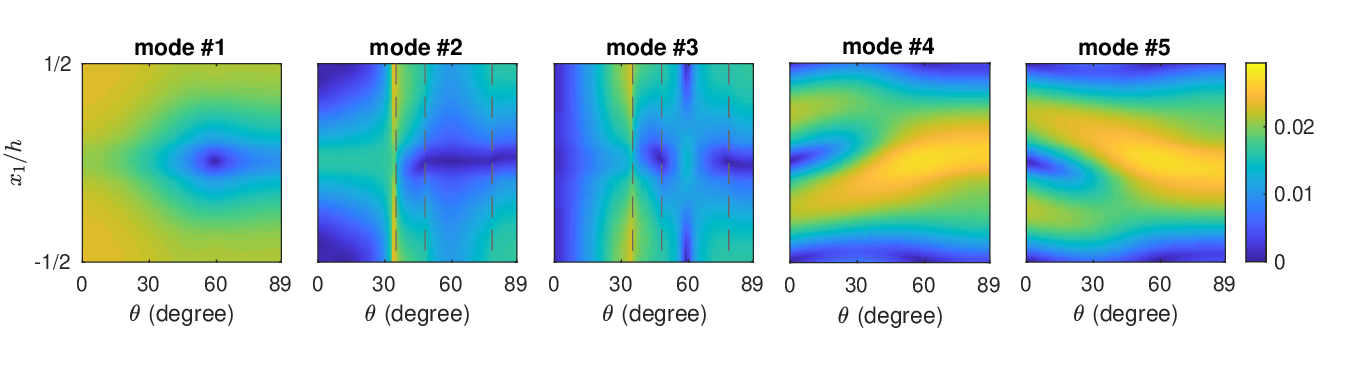}
		\caption{$\bar{u}_1$ amplitudes\label{fig:msu1abs}}
	\end{subfigure}
	\begin{subfigure}[b]{1\linewidth}
		\centering\includegraphics[width=0.92\linewidth]{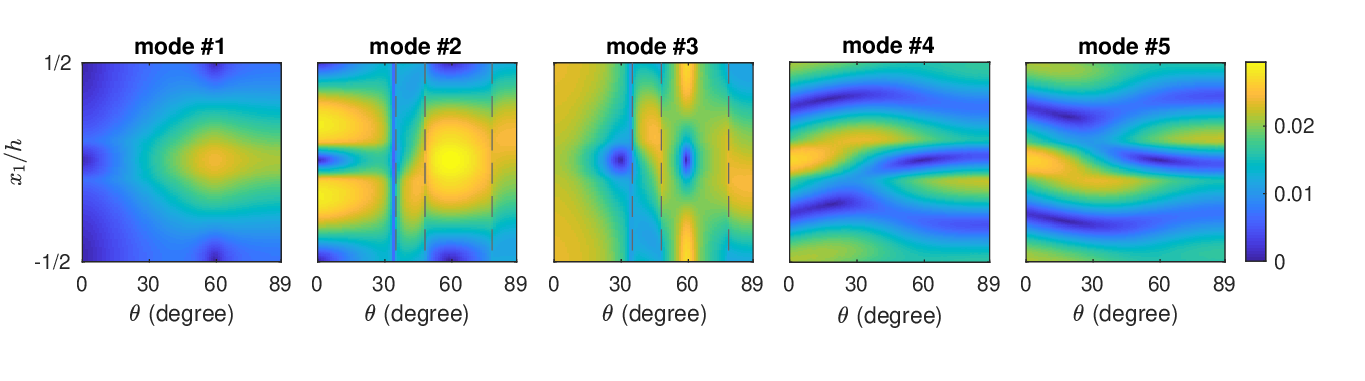}
		\caption{$\bar{u}_2$ amplitudes\label{fig:msu2abs}}
	\end{subfigure}
	\begin{subfigure}[b]{1\linewidth}
		\centering\includegraphics[width=0.92\linewidth]{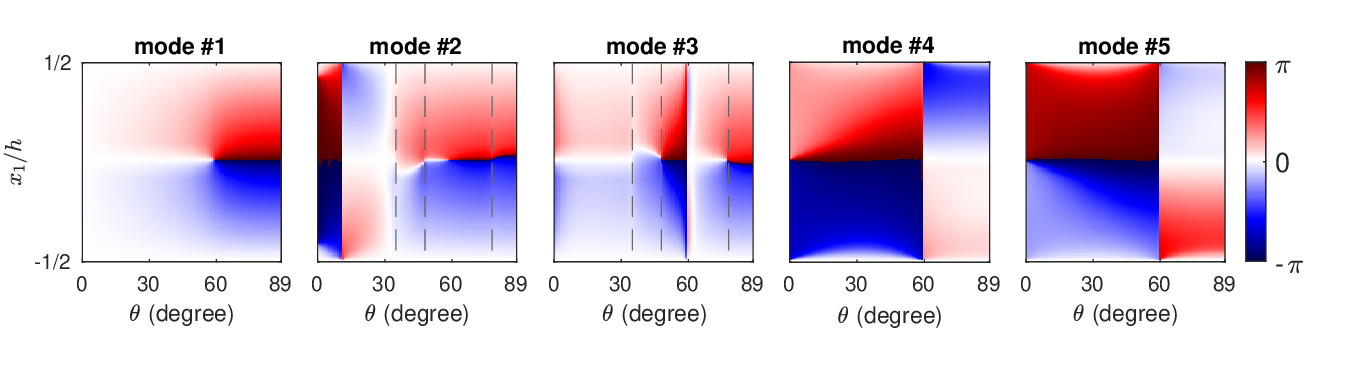}
		\caption{$\bar{u}_1$ arguments\label{fig:msu1arg}}
	\end{subfigure}
	\begin{subfigure}[b]{1\linewidth}
		\centering\includegraphics[width=0.92\linewidth]{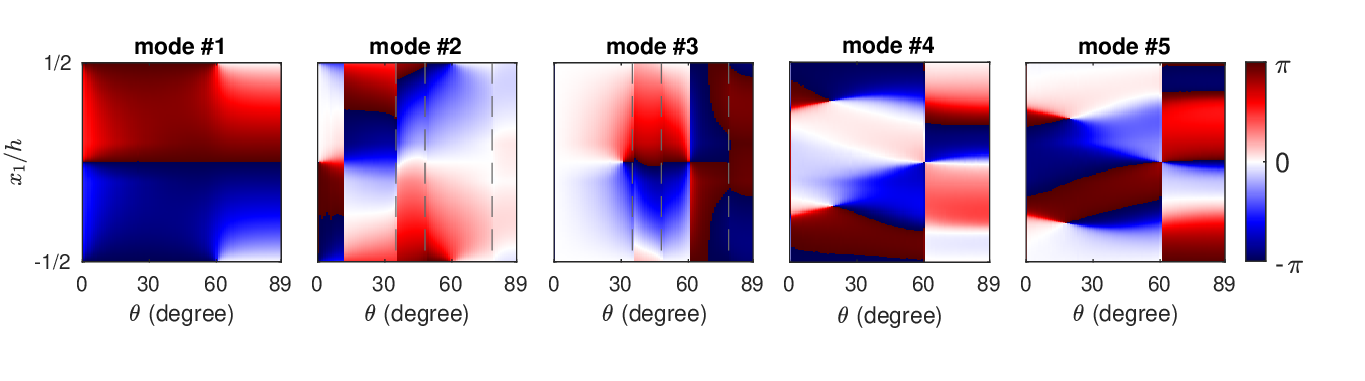}
		\caption{$\bar{u}_2$ arguments\label{fig:msu2arg}}
	\end{subfigure}
	\caption{Normalized displacement mode shapes of the PC evaluated along the ~$x_1$ interface as functions of incident angle~$\theta$:~(\subref{fig:msu1abs}) amplitudes and (\subref{fig:msu1arg}) arguments of~$\bar{u}_1$,~(\subref{fig:msu2abs}) amplitudes and (\subref{fig:msu2arg}) arguments of~$\bar{u}_2$. The angles associated with EP and CAs are marked by the vertical dashed lines in modes 2 and 3 graphs. \label{fig:dispms} }
\end{figure}

In the oblique scattering problem, the reflected/transmitted waves in two domains are composed of all the eigen-solutions at the corresponding~$k_1h$ value of the incident wave. Therefore, the physical properties of scattering signals are affected by not only the~$k_2$ eigenvalues but also the mode shapes. \Cref{fig:dispms} shows the displacement mode shapes of the periodic cell along the interface as functions of incident angle. The presented five mode shapes are the displacement mode shapes corresponding to the five modes in \cref{fig:transband}. The unit cell inversion center is set to be~$x_1=0$. For each mode and each angle~$\theta$, the complex displacements are normalized so that~$\norm{[\vec{\bar{u}}^{(m)}_1,\ \vec{\bar{u}}^{(m)}_2]^\top}=1$, and~$\Im [\bar{u}^{(m)}_{1}(x_1=-h/2)]=0$, where the superscript denotes the~$m$-th mode, and the bold symbol denotes the vector form of a quantity along~$x_1$. \Crefrange{fig:msu1abs}{fig:msu2abs} are the amplitudes of displacements in~$x_1$ and~$x_2$ directions. \Crefrange{fig:msu1arg}{fig:msu2arg} are the complex arguments of displacements in~$x_1$ and~$x_2$ directions. For mode 2 and 3, three vertical dashed lines are plotted at EA (35.1$^\circ$), CA3 (48.8$^\circ$), and CA4 (78.4$^\circ$).

It can be seen that, for modes 1, 4, and 5 the displacements~$\vec{\bar{u}}_{1,2}^{(1,4,5)}$ are relatively smooth and continuous with respect to the variation in angle, as their eigenvalues~$k_2^{(1,4,5)}(\theta)$ do not undergo branch crossing or coalescence. For mode 2 and 3, clear transitions in~$\vec{\bar{u}}_{1,2}^{(2,3)}$ can be found at the three special angles marked by the dashed lines, both in their amplitudes and the arguments. Spectral transitions in the~$k_2$ eigenvalues therefore lead to drastic changes in the associated wave mode shapes. Although only the displacement part of the mode shapes are shown, the stress components also share similar transitions at these angles.

The mode shape patterns change their symmetries as well when the incident angle sweeps through the EA and CAs. It turns out that the EP and CAs have underlying relations with spontaneous symmetry breaking. The PC unit cell shown in \cref{fig:setup} possesses parity symmetry with respect to its inversion center~$x_1=0$. The governing equation is invariant if the PC's parity is reversed. Along a line of constant~$x_2$, the displacement wave associated with a certain mode with a positive~$k_1$ value is
\begin{equation}
    u_{1,2}^+ (x_1,t)=\abs{\bar{u}_{1,2}(x_1)}\exp\Big(\im (\bar{\varphi}_{1,2}(x_1)+\bar{\varphi}_0+k_1x_1 -\omega t)\Big),
\end{equation}
where the superscript~$+$ denotes a wave with positive~$k_1$ value,~$\bar{\varphi}(x_1)$ is the complex argument of~$\bar{u}_{1,2}(x_1)$, and the~$k_2$ dependence is omitted since~$x_2$ is constant here. Here~$\bar{\varphi}_0$ is an arbitrary real phase applied to the eigen-mode displacement field, and should be consistent for both~$u_1$ and~$u_2$. Similarly, for the same wave at~$-x_1$, we have
\begin{equation}
    u_{1,2}^+ (-x_1,t)=\abs{ \bar{u}_{1,2}(-x_1)}\exp\Big(\im (\bar{\varphi}_{1,2}(-x_1)+\bar{\varphi}_0-k_1x_1 -\omega t)\Big).
\end{equation}
Now we consider a wave of the same mode propagating in a reversed direction, the wavevector component~$k_1$ becomes negative, and the displacement at~$-x_1$ is
\begin{equation}\label{eq:back-x1}
    u_{1,2}^- (-x_1,t)=\abs{ \bar{u}_{1,2}(-x_1)} \exp\Big(\im (-\bar{\varphi}_{1,2}(-x_1)-\bar{\varphi}_0+k_1x_1 -\omega t)\Big).
\end{equation}
\Cref{eq:back-x1} is in such a form because~$\Re [ u_{1,2}^+ (t)]=\Re [ u_{1,2}^- (-t)]$ must be satisfied. Given the parity symmetry of the unit cell, it is expected that
\begin{equation}\label{eq:symme}
    u_{1,2}^+ (x_1,t)=u_{1,2}^- (-x_1,t)
\end{equation}
for a mode in the symmetry unbroken phase. \Cref{eq:symme} reveals that a symmetry-unbroken mode shape (for the~$m$-th mode) must satisfy:
\begin{align} \label{eq: intrasym}
    \begin{split}
        &\abs{ \bar{u}_{1,2}^{(m)}(x_1)}=\abs{ \bar{u}_{1,2}^{(m)}(-x_1)},\\
        &\bar{\varphi}_{1,2}^{(m)}(x_1)+\bar{\varphi}_{1,2}^{(m)}(-x_1)=-2\bar{\varphi}^{(m)}_0=\mathrm{const.} 
    \end{split}
\end{align}
Since the displacement vector is normalized in such a way that~$\bar{\varphi}^{(m)}_1(-h/2)=0$, we have~$\bar{\varphi}^{(m)}_0=0$. In other words, the amplitudes must be symmetric with respect to the cell inversion center while the arguments must be anti-symmetric.
%\Cref{eq: intrasym} applies as well to stress mode shapes. 
It is shown in \cref{fig:dispms} that the symmetry unbroken conditions in \Cref{eq: intrasym} are satisfied for the branches with purely real~$k_2$ eigenvalues, i.e., mode 1 from~$0^\circ$ to~$89^\circ$, mode 2 from CA3$=48.8^\circ$ to~CA4$=78.4^\circ$, and mode 3 from~$0^\circ$ to EA and from CA3 to CA4. 
For the branches with complex eigenvalues (e.g., mode 4 and 5, and partially mode 2 and 3), the corresponding modes are in symmetry broken phases. The spontaneous symmetry breaking occurs in reflected modes as well. Between the two CAs, the reflected mode 2 has imaginary~$k_2$ and displacement mode shapes that violate \cref{eq: intrasym}. For all the branches with broken modal symmetry, their averaged energy fluxes at the interface will be zero.

\begin{figure}[h!]
	\begin{subfigure}[b]{0.47\linewidth}
		\centering\includegraphics[height=120pt]{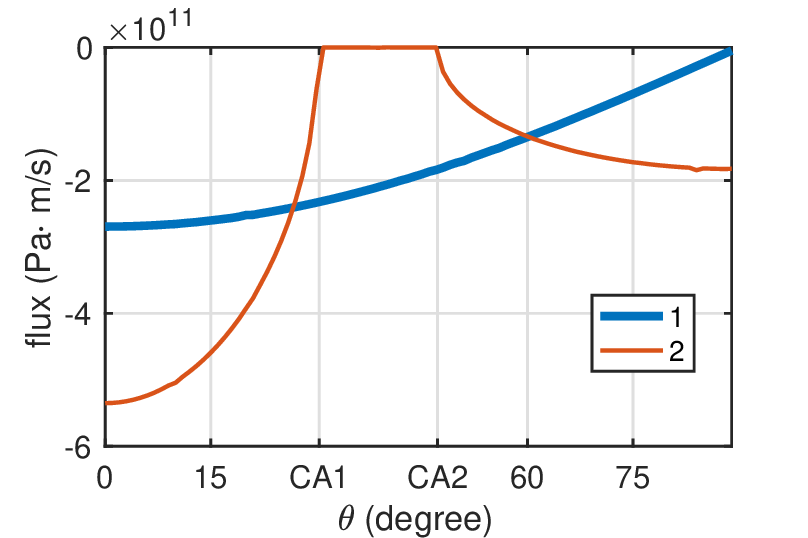}
		\caption{\label{fig:refflux}}
	\end{subfigure}%
	\begin{subfigure}[b]{0.47\linewidth}
		\centering\includegraphics[height=120pt]{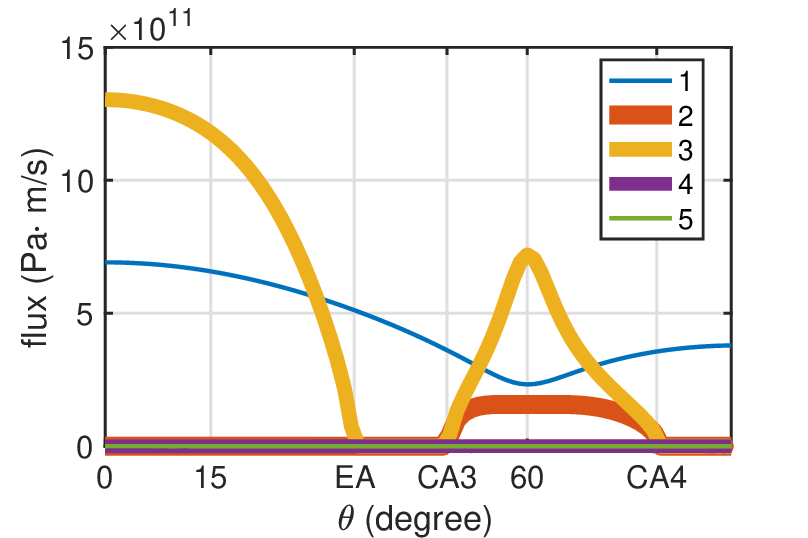}
		\caption{\label{fig:transflux}}
	\end{subfigure}
	\caption{Time and unit cell averaged modal fluxes in~$x_2$ direction:~(\subref{fig:refflux}) modes of the homogeneous medium and~(\subref{fig:transflux}) modes of the PC.\label{fig:flux} }
\end{figure}

%The properties of each modes can also be analyzed through their energy fluxes. 

For the~$m$-th mode, its energy flux at the the interface can be calculated as
\begin{equation}
    F^{(m)}_2 (x_1)=\int_{-h/2}^{h/2}-\frac{1}{2}\Re\left[\bar{\sigma}^{(m)}_{2j}(x_1) \cdot (\partial_t \bar{u}^{(m)}_{j}(x_1))^*\right]\mathrm{d} x_1,
\end{equation}
where the summation is done over~$j\in\{1,2\}$, and superscript~$*$ denotes complex conjugate. The flux represents the time-averaged stress wave intensity in~$x_2$ direction. \Crefrange{fig:refflux}{fig:transflux} show the fluxes of the homogeneous medium modes and the PC modes, related to the bands in \cref{fig:band}. Each mode shape used for this plot is normalized so that $\norm{[\vec{\bar{u}}^{(m)}_1,\ \vec{\bar{u}}^{(m)}_2]^\top}=1$. In comparison with \cref{fig:band}, one finds that an eigenvalue branch with complex-valued~$k_2$ (see \cref{fig:band}) will have zero flux at the homogeneous-PC interface, e.g., reflected mode 2 between CA1 and CA2 in \cref{fig:refflux} and transmitted mode 2 between~$0^\circ$ and CA3 in \cref{fig:transflux}. This is expected since the modes with complex~$k_2$ are evanescent in~$x_2$ and only propagate along the surface ($x_1$ interface). For the modes that possess branch points, the first order derivatives of their fluxes with respect to~$\theta$ becomes discontinuous at the special angles CAs and the EA. Therefore these modes (mode 2 of the homogeneous medium, mode 2 and 3 of the PC) undergo phase transitions at the CAs and the EA. On the other hand, a branch with real-valued~$k_1$ is capable of transporting energy in~$x_2$ and its flux is non zero. The branches with zero net flux are exactly the ones in the symmetry broken phase.

\begin{figure}[!ht]
\centering
	\begin{subfigure}[b]{0.5\linewidth}
		\centering\includegraphics[width=0.8\linewidth]{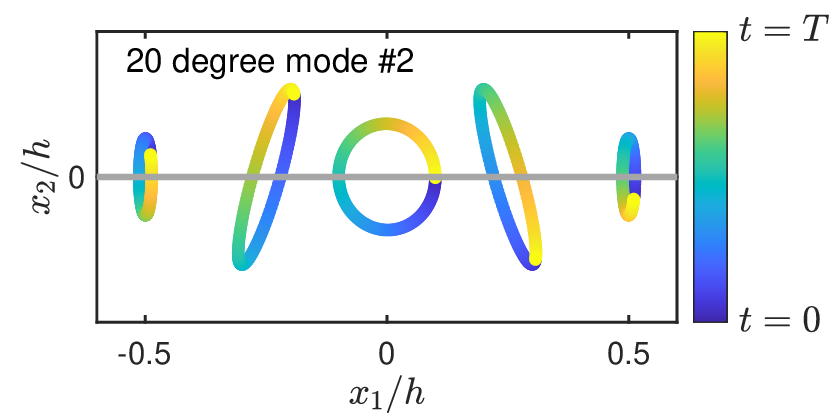}
		\caption{imaginary~$k_2$\label{fig:m2_20}}
	\end{subfigure}%
	\begin{subfigure}[b]{0.5\linewidth}
		\centering\includegraphics[width=0.8\linewidth]{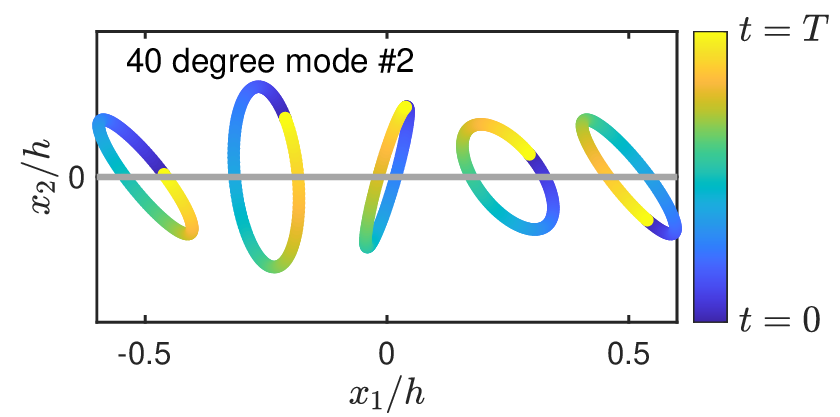}
		\caption{complex~$k_2$\label{fig:m2_40}}
	\end{subfigure}
	\begin{subfigure}[b]{0.5\linewidth}
		\centering\includegraphics[width=0.8\linewidth]{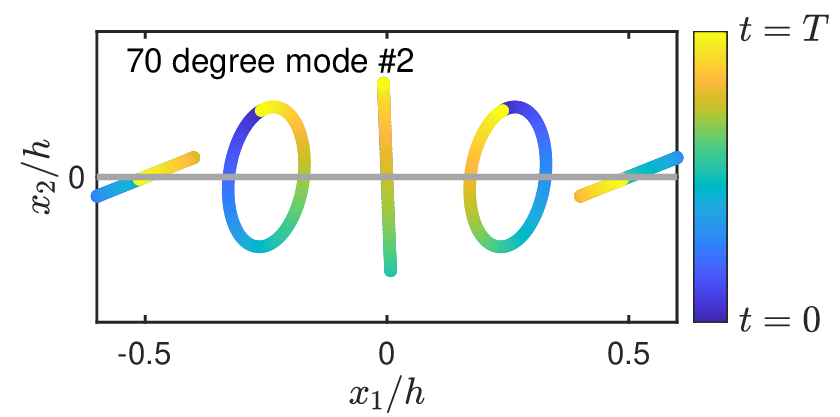}
		\caption{real~$k_2$\label{fig:m2_70}}
	\end{subfigure}
	\caption{Examples of particle trajectories. \label{fig:mode2ellipse} }
\end{figure}

% ellipse study
The breaking of symmetry affects the particle motion trajectories as well. In time domain, the shape of the particle deformation given by~$\Re[u_{1} (x_1,t),u_{2} (x_1,t)]$ is in general an ellipse or a circle. Special case occurs when~$\abs{\angle u_1-\angle u_2}\in\{0,\pi\}$ and the motion will be polarized as a straight line. Furthermore, the handedness of the the trajectory is determined by the phase difference between~$u_1$ and~$u_2$. The particle moves clockwise in time if~$\angle(u_1/u_2)<0$. The handedness becomes counterclockwise if~$\angle(u_1/u_2)>0$. 
Based on the periodicity~$\angle (u_1/u_2) (x_1) =\angle (u_1/u_2) (x_1+h)$ and the parity symmetry \cref{eq: intrasym}, it can be seen that, only when~$k_2$ is purely real, the motions at cell inversion centers~$x_1=0,\pm h/2$ will be polarized as lines, and the motions at~$x_1$ and~$-x_1$ must have reversed handedness.
Three examples are shown in \cref{fig:mode2ellipse}. The particle trajectories are plotted for five points uniformly sampled along~$x_1$. The lack of symmetry for complex $k_2$ is evident here. Even for a purely imaginary $k_2$ the behavior is distinct from that of a purely real $k_2$.
%In \cref{fig:m2_20}, the~$k_2$ eigenvalue is purely imaginary.

\begin{figure}[!ht]
\centering
		\centering\includegraphics[width=0.38\linewidth]{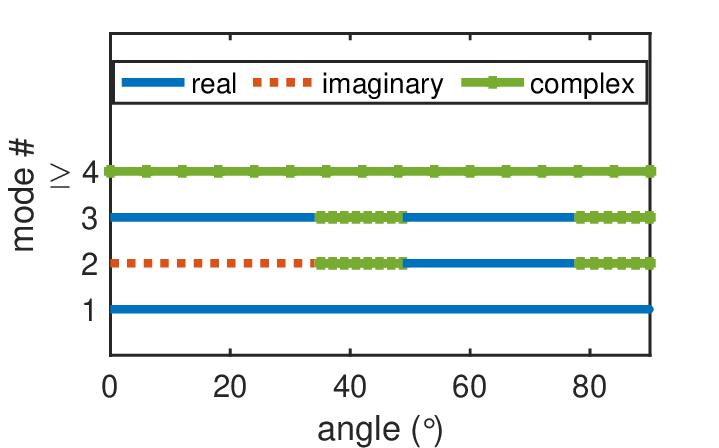}
	\caption{Summary of~$k_2$ eigenvalue types for the PC eigen-modes. \label{fig:k2kind} }
\end{figure}

% summary in table
\begin{table}[!ht]
\centering
\begin{tabular}{|l|l|l|l|}
\hline
$k_2$     & real    & imaginary & complex  \\ 
\hline
eigen-mode symmetry & unbroken & broken & broken \\
\hline
flux & non-zero & zero & zero\\
\hline
trajectory at inversion center & line & ellipse & ellipse \\
\hline
geometry$^a$ 
& anti-symmetric & symmetric & asymmetric\\
\hline
chirality$^b$ & anti-symmetric & symmetric & asymmetric\\
\hline
\end{tabular}
\caption{Modal properties for different types of~$k_2$ eigenvalues. $^a$ Geometry: refers to the shapes of the trajectories w.r.t. RUC inversion center. $^b$ Chirality: refers to the handedness directions of the trajectories w.r.t. RUC inversion center.
}\label{table:1}
\end{table}

To summarize, the critical angles (CAs) and the exceptional points (EPs) are identified as the spectral branch points between the complex and real eigenvalue branches. As the incident angle passes across these limits (CAs and EAs), certain eigen-mode(s) will switch between propagating ones and evanescent ones. A propagating mode carries energy in~$x_2$ while an evanescent one only propagates along the~$x_1$ interface. At the branch points, spontaneous symmetry breaking occur in the corresponding mode shapes. The modal energy fluxes will vanish in the symmetry broken phases. The geometry and handedness of the motion trajectories are highly dependent on the~$k_2$ eigenvalues as well. The types of~$k_2$ eigenvalues of the PC modes are summarized in~\cref{fig:k2kind}. The angle-dependent modal behaviors, summarized in~\cref{table:1}, will cause qualitative changes on the scattering signal, and will be used as the base of the proposed sensing application.

\section{Scattering analysis}\label{sec:signal}
In a scattering configuration (with far-field assumption), the displacement and stress fields may be written as weighted summations of all the potential mode shapes (the lowest order of which are shown in \cref{eq:eig}). 
On the transmission side of the interface (PC domain) where the measurements are generally made, the field quantities are:
\begin{align}
\begin{split}
    u_{1,2}(\theta,x_1,x_2)&=\Sigma_n T^{(n)} \bar{u}_{1,2}^{(n)}\exp \left[ \im (k_1(\theta)x_1+k_2^{(n)} x_2)\right],\\
      \sigma_{ij}(\theta,x_1,x_2)&=\Sigma_n T^{(n)} \bar{\sigma}_{ij}^{(n)}\exp \left[ \im (k_1(\theta)x_1+k_2^{(n)} x_2)\right],
    \end{split}
\end{align}
where~$T^{(n)}$ is the coefficient of the~$n$-th transmitted mode and can be determined in a number of ways including one based on Betti-Rayleigh reciprocity\cite{Mokhtari2020a}.
The first three modes turn out to have dominant weights in determining the scattering response, even though for portions of the angular spectrum, modes 2 and 3 are in fact evanescent. The evanescent modes are required to satisfy the continuity conditions at the interface, therefore the scattered field associated with different angles of arrival are expected to lie in the subspace~$M(\theta)$ spanned by these dominant modes. 
Assuming that~$n_p$ measurements are taken per unit cell along the interface, and~$n_c$ cells are used, we have
\begin{equation}
    \vec{s}(\theta)\in M(\theta)=\mathrm{span}\{ \vec{m}^{(1)} (\theta),\vec{m}^{(2)} (\theta),\vec{m}^{(3)} (\theta) \}\subset\mathbb{C}^{n_p n_c}
\end{equation}
where~$\vec{s}$ is the frequency domain complex amplitudes measured at the sensors with  length~$n_pn_c$, and~$\vec{m}^{(n)}$ is the~$n$-th mode shape vector (stress or displacement quantities) of the same length sampled at the same locations.

\section{Source localization using deep learning of the modes}\label{sec:sensing}

It was concluded in the previous section that the PC mode shapes vary drastically for different angles of incidence due to the existence of the branch points. This is most pronounced in their symmetry properties. Therefore, both the scattered field and the underlying subspace~$M(\theta)$ will inherently possess high variance with respect to the angles. 
The proposed source localization strategy is stated as follows. For any complex vector~$\vec{s}$ taken from measurement, if it is identified as an element of subspace~$M(\theta)$, then the incident angle is~$\theta$. This method takes advantage of the fact that the subspace spanned by the dominant modes are sensitive to incident angles. To implement the localization, artificial neural networks (ANNs) can be used to learn and extract the abstract features of the subspace~$M(\theta)$. In the following, an example of source localization is shown using a feed-forward deep learning (DL) neural network (NN) setup. Notice that other NN architectures such as convolutional neural networks (CNNs) can also be used for this purpose.

\subsection{Data preprocessing}

The quantity selected for measurement in this example is the normal stress component~$\sigma_{11}$. In the example here,~$n_p=6$ measurements are taken per unit cell, and~$n_c=10$ cells are used in total. The data collection points are measured along the interface~$x_2=0$ with a uniform spacing of~$h/6$. Such a sensor array is referred to as a uniform linear array (ULA).

The input to the ANN will represent complex vectors in~$M(\theta)$ in the training stage, and the output will indicate the angle~$\theta$. 
For each incident angle~$\theta$, 2100 training samples are prepared and labelled by the associated integer angle~$\theta\in[1^\circ,89^\circ]$. Each training sample is given by
\begin{equation}
    \vec{t}(\theta)=c^{(1)}\vec{m}^{(1)} (\theta)+c^{(2)}\vec{m}^{(2)} (\theta)+c^{(3)}\vec{m}^{(3)} (\theta).
\end{equation}
Here,~$\vec{m}^{(n)} (\theta)$ is the~$n$-th~$\sigma_{11}$ mode shape vector of length~$n_p n_c$ associated with incident angle~$\theta$, and it includes the~$\im k_1(\theta) x_1$ phase. The complex coefficient~$c^{(n)}$ of each mode is randomly selected in such a way that both~$\Re c^{(n)}$ and~$\Im c^{(n)}$ have uniform distributions between -0.5 and 0.5. 
A validation set is prepared in the same fashion and consists of 450 labelled samples~$\vec{v}(\theta)$ for each value of $\theta$. Unlike the training and validation sets, the test set consists of the scattering signals~$\vec{s}$ instead of random vectors in~$M(\theta)$. It should be highlighted that the test set is unknown to the NN and will therefore provide the evaluation of the sensing performance.

It is necessary to normalize input data properly in order to render it independent of source strength. The complex vectors~$\vec{b}=\vec{t},\ \vec{v},\ \mathrm{or}\ \vec{s}$ are normalized so that
\begin{equation}
    \norm{\vec{b}}=1.
\end{equation}
Then, the complex vectors~$\vec{b}=\vec{t},\ \vec{v},\ \mathrm{or}\ \vec{s}$ are converted into real-valued arrays~$\vec{b}'$ before feeding into the ANN:
\begin{equation}\label{eq:comarray}
    \vec{b}'=\begin{pmatrix}
    \abs{\vec{b}}\\
    \cos\angle\vec{b}\\
    \sin\angle\vec{b}
    \end{pmatrix}\in\mathbb{R}^{3n_p n_c}.
\end{equation}
The amplitude and angle operators are applied to each component of vector $\vec{b}$ separately. The redundancy of applying the $\sin$ and $\cos$ operators separately is intentional.

\subsection{Neural network}

\begin{figure}[!ht]
\centering
		\centering\includegraphics[width=0.4\linewidth]{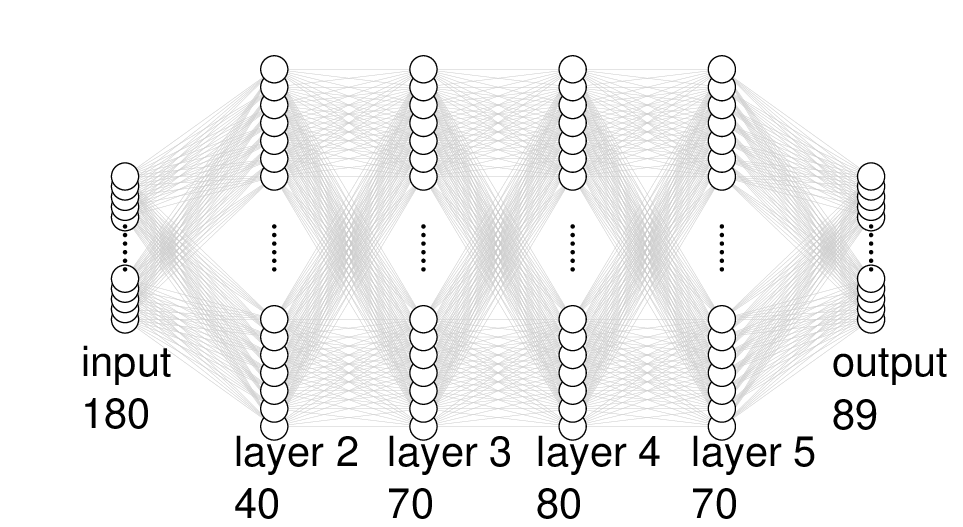}
	\caption{The feed-forward (fully-connected) neural network architecture. The numbers of neurons (excluding the bias unit) are labelled under the layers. The bias units of the first five layers are not shown. \label{fig:NNarch} }
\end{figure}

In this example, a feed-forward NN is used for deep learning of the eigen-modes and for classifying the incident angle of unknown signals. The NN architecture is shown in~\cref{fig:NNarch}. It includes an input layer with 180 neurons for $n_p = 6$ and $n_c = 10$. 
The input array is formatted based on~\cref{eq:comarray}. The output layer has 89 neurons with each representing an integer angle~$\in[1^\circ,89^\circ]$ through one-hot encoding. The angle labels of the training and validation samples are represented by binary vectors of length 89.

At the hidden layers (2 to 5), the NN operation is as follows. Let~$s_j$ denote the number of neurons (bias unit not included) of the~$j$-th layer. The neuron values at the~$j$-th layer can be written as a column vector~$\vec{a}^{(j)}$. The vector~$\vec{a}^{(j)}$ has a length of~$s_j+1$ to include the bias unit (for~$j\leq5$). For example, the training input is~$\vec{a}^{(1)}=[1,\vec{t}']^\top$, and~$\vec{t}'$ is a real vector of length 180. The value of each neuron is obtained by first computing a weighted sum of all neurons (with the bias unit) in the previous layer. This can be written as a matrix calculation:
\begin{equation}
    \vec{z}^{(j)}=\Theta^{(j-1)} \vec{a}^{(j-1)}
\end{equation}
where~$\Theta^{(j-1)}$ is a~$s_j$ by~$s_{j-1}+1$ real matrix representing the connection weights between layer~$j-1$ and layer~$j$. Then, the batch normalizing transform is applied to the updated vector~$\vec{z}^{(j)}$ to obtain~$\bar{\vec{z}}^{(j)}$. The batch normalization effectively re-centers and re-scales the data array to achieve faster and more stable performance of the NN\cite{Ioffe}. This is followed by a non-linear activation using the rectified linear unit (ReLU) function:
\begin{equation}
    \mathrm{ReLU}(x)=\mathrm{max}(0,x).
\end{equation}
The ReLU activation is widely chosen for ANNs due to its multiple advantages\cite{Glorot} such as better gradient propagation and computational efficiency.
Finally the updated state of the layer is given by
\begin{equation}
     \vec{a}^{(j)}=\mathrm{ReLU}(\bar{\vec{z}}^{(j)}).
\end{equation}

For the output layer, we have
\begin{equation}
    \vec{z}^{(6)}=\Theta^{(5)} \vec{a}^{(5)}.
\end{equation}
The final output is activated through the sigmoid function:
\begin{equation}
    \vec{a}^{(6)}=S(\vec{z}^{(6)})=\frac{1}{1+\exp(-\vec{z}^{(6)})}.
\end{equation}
The sigmoid function, for each output neuron, returns a value in the range 0 to 1. The output~$\vec{h}=\vec{a}^{(6)}$ predicts the probability of each angle.

\begin{figure}[!ht]
\centering
		\centering\includegraphics[width=0.37\linewidth]{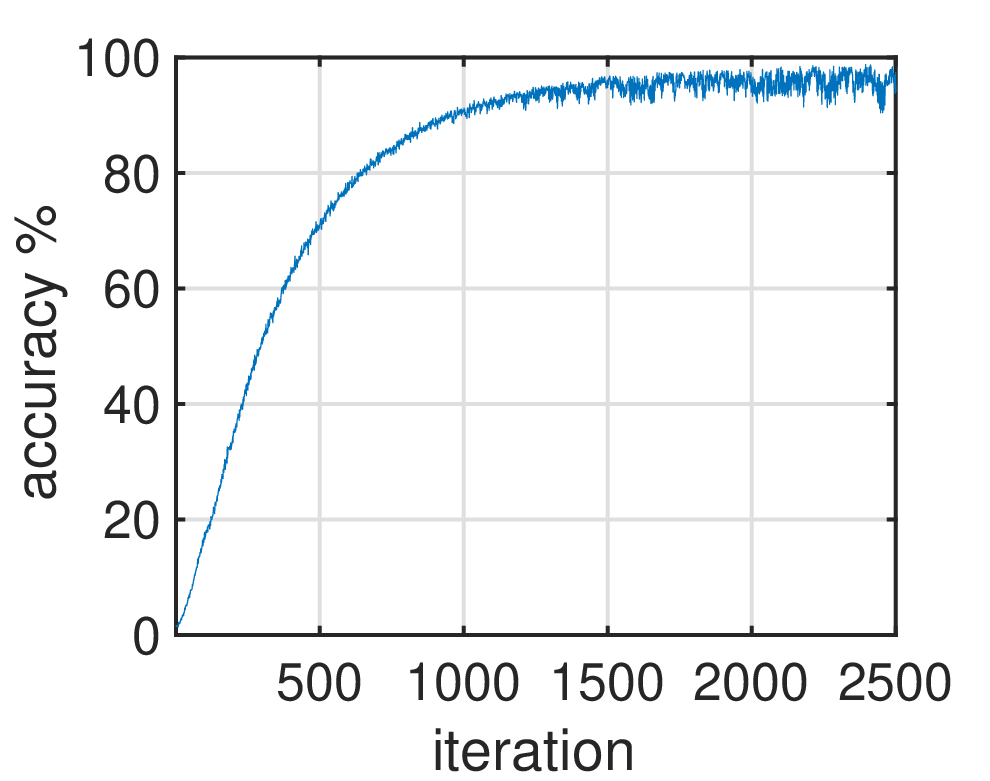}
	\caption{The convergence plot shows the training accuracy for each iteration (mini-batch). 
		\label{fig:train} }
\end{figure}

The NN is first randomly initialized. In the training process, the training set (2100 samples per angle) is randomly separated into 25 mini-batches and fed into the NN for 100 epochs. At each training iteration, one batch of training data is passed through the NN. A cross-entropy cost function\cite{kevin} is then evaluated as
\begin{equation}
    J(\vec{P})=-\frac{1}{N} \sum^{N}_{n=1} \sum^K_{i=1} \left(T_{i,n}\log(h_{i,n}) +(1-T_{i,n}) \log(1-h_{i,n})\right),
\end{equation}
where~$N=2100*89/25$ and~$K=89$ are the numbers of samples and incident angles, respectively. 
The target value of~$i$-th output neuron for the~$n$-th sample is~$T_{i,n}\in\{0,1\}$. The output value of~$i$-th neuron for the~$n$-th sample is~$h_{i,n}\in[0,1]$. Here, the vectorized variable~$\vec{P}$ contains all the NN parameters to be optimized, including the connection weights~$\Theta$, the off-set factors, and the re-scale factors. At the end of each iteration, the cost~$J$ and its gradients with respect to~$\vec{P}$ are evaluated. Then, the parameters~$\vec{P}$ are updated based on gradient descent and will be used for the next iteration. The training process aims to minimize the cost and find the best set of NN parameters. \Cref{fig:train} shows the convergence of the training accuracy for each iteration. 

The final training accuracy for the last batch is 93.93\%. Then the NN is tested using a validation set that has~450 labelled samples per angle. The validation set is a secondary set of data that is not learned by the NN and can therefore provide an unbiased evaluation of the NN fitness. In this case, the accuracy for the validation set is~$93.96\%$. Recall that all the samples in the training set and the validation set are made up by randomly weighted eigen-modes. The relatively good accuracy achieved on these eigen-modes indicates that the subspace spanned by the eigen-modes indeed has angle-sensitive features that can be learned by the NN.

\subsection{Results and discussion}

\begin{figure}[!ht]
\centering
	\begin{subfigure}[b]{0.49\linewidth}
		\centering\includegraphics[width=0.75\linewidth]{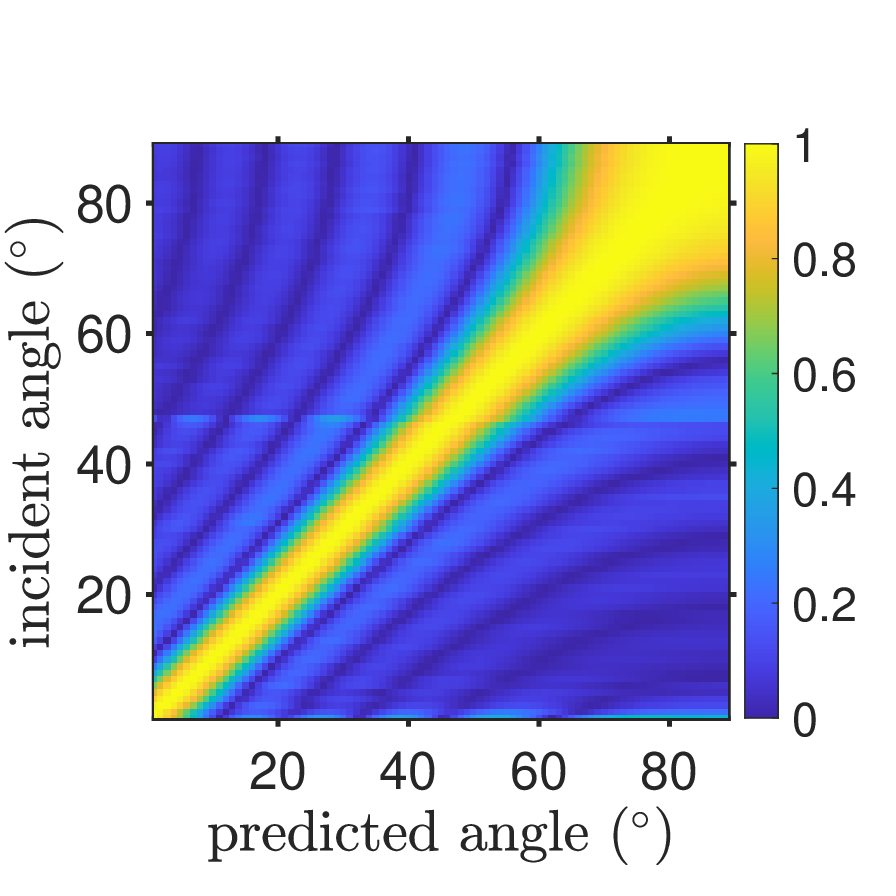}
		\caption{DAS: 84.27\% accuracy \label{fig:das25}}
	\end{subfigure}%
	\begin{subfigure}[b]{0.49\linewidth}
		\centering\includegraphics[width=0.75\linewidth]{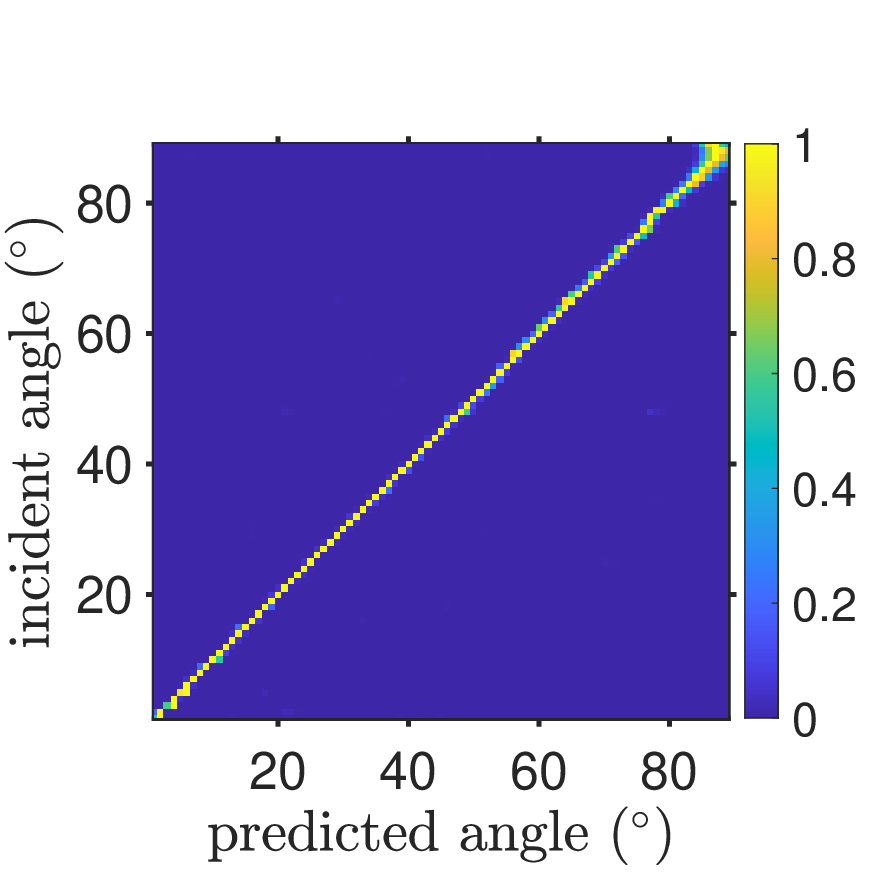}
		\caption{DNN: 93.26\% accuracy \label{fig:dnn25}}
	\end{subfigure}\\
		\begin{subfigure}[b]{0.49\linewidth}
		\centering\includegraphics[width=0.8\linewidth]{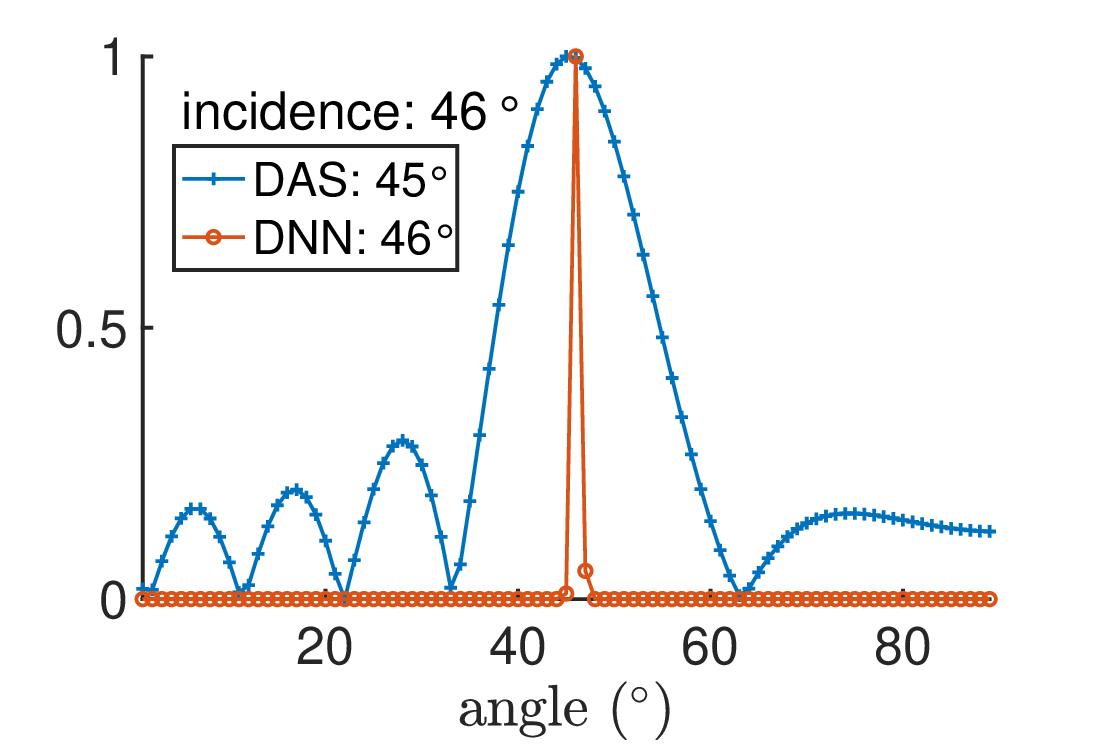}
		\caption{$46^\circ$ incidence \label{fig:comp46}}
	\end{subfigure}%
	\begin{subfigure}[b]{0.49\linewidth}
		\centering\includegraphics[width=0.8\linewidth]{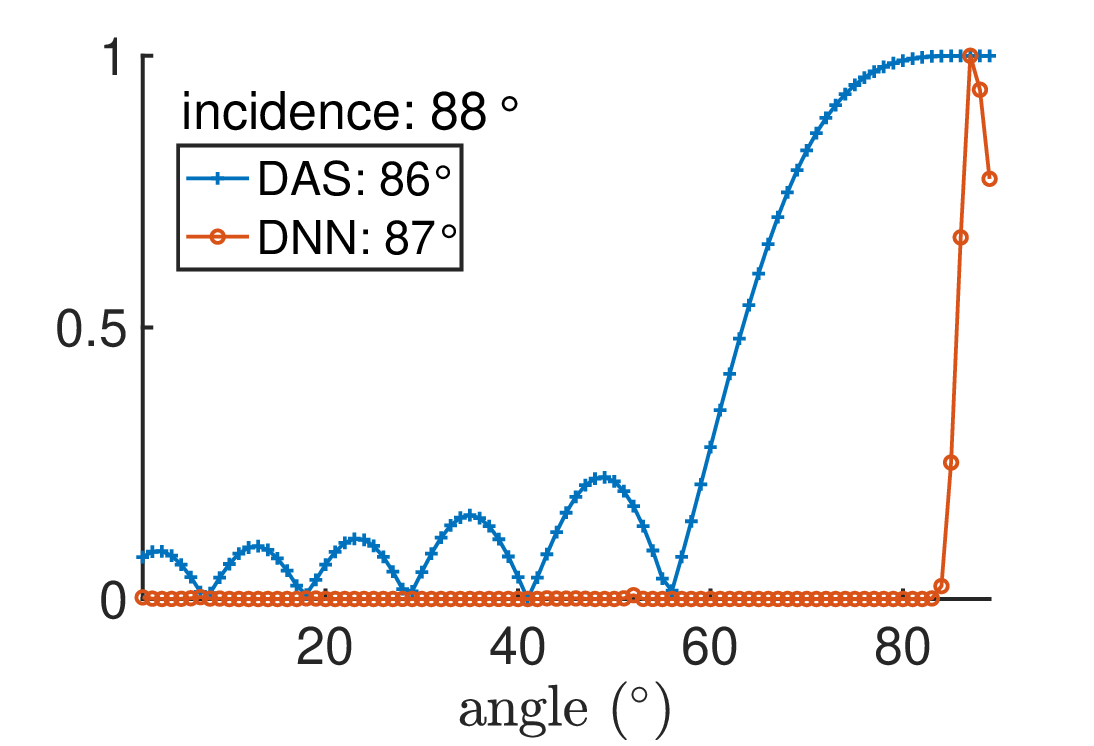}
		\caption{$88^\circ$ incidence \label{fig:comp88}}
	\end{subfigure}\\
	\caption{Localization results of DAS and DNN. To compare between two methods, all the outputs are re-scaled from 0 to 1 and are shown in the linear scale. \label{fig:rescomp} }
\end{figure}

The test set which consists of the scattering signals from~$1^\circ$ to~$89^\circ$ (calculated in this case based on Betti-Rayleigh reciprocity principle\cite{Mokhtari2020a}) is fed into the DNN to verify the sensing performance. The signals used here are slightly corrupted with additive white Gaussian noises added to the frequency domain complex amplitude.
The signal-to-noise-ratio (SNR) is~$25$ dB. 
The localization performance of the trained DNN is compared with a conventional method, delay-and-sum (DAS). A simplified DAS algorithm is used here and is shown in the appendix. Detailed studies of DAS can be found in literature\cite{Bai2013,Perrot2021}. \Cref{fig:rescomp} shows the performance comparison between DAS and DNN. The DAS~(\cref{fig:das25}) produces wider main lobes and has higher side lobes. In contrast, the DNN provides (\cref{fig:dnn25}) sharp peaks at the predicted angles, and the side lobe levels are significantly suppressed. The DNN also leads to a higher accuracy (93.26\%) in identifying the incident angles compared with the DAS (84.27\%). A detailed example can be seen in~\cref{fig:comp46}, where the DNN outperforms the DAS and leads to the correct angle of incidence. However, the trained DNN is not fully error-free either. In~\cref{fig:comp88} for instance, the DNN mistakenly predicts the~$88^\circ$ angle to be~$87^\circ$. The end-fire sources present challenges for any AoA estimation algorithm. Improved performance may be achievable by adjusting the NN architecture and fine tuning of the hyper parameters, or more fundamentally by design of micro-structure for higher modal sensitivity at such angles of incidence. In general, the proposed approach using the NN to identify angle of incidence shows strong potential and benefits in the localization application. This approach can be easily implemented as a parallel and complementary procedure with traditional methods.

\section{Conclusion}
In this work, we exploit the eigen-wavevector band structure of micro-structured media under oblique scattering and present the sensing potentials based on deep learning of the angle-dependent modal features. In the studied oblique stress wave scattering problem, the modal symmetry breaking at the critical angles and the exceptional point is identified and discussed in detail through their modal shape symmetry, flux, and polarization. It is understood that the scattering signals lie predominantly in the subspace spanned by the lower eigen-modes and the subspace has inherently strong dependence on the incident angle. An artificial neural network is trained with random sampling of the subspace spanned by the eigen-modes as the training input. The trained neural network is able to identify the angle-dependent features of the modal subspace and shows major improvements in identifying the angle of incidence based on scattering as input test data in comparison with standard delay-and-sum approach.

To summarize, the proposed method:
\begin{itemize}
  \item highlights and utilizes the physics of the band structure and eigen-modes for sensing purposes;
  \item shows that artificial neural networks can be trained using eigen-modes in order to identify incident angles;
  \item can be continuously improved  with the fast-growing deep learning techniques,
  \item can be optimized through careful micro-structural design leading to feature-rich and wavevector sensitive modal subspace.
\end{itemize}
The scope of this paper is limited to computational investigation, and the example shown is a theoretical proof of concept. Future work should include: (1) developing a physics-informed deep neural network for more complicated scattering scenarios, e.g., multi-source and multi-frequency; (2) topology optimization of the micro-structured medium; and (3) experimental verification of the proposed sensing approach.

\begin{acknowledgments}
The authors acknowledge NSF support for this work through grant \#1825969 to University of Massachusetts, Lowell, and grant \#1825354 to Illinois Institute of Technology, Chicago.
\end{acknowledgments}

\noindent\textbf{DATA AVAILABILITY}
\newline
The data that support the findings of this study are available from the corresponding author
upon reasonable request.

\appendix

\section{Delay-and-sum}\label{app:das}
Delay-and-sum (DAS) is one of the fundamental source localization algorithms. It utilizes the spatial discrete Fourier transform to identify the dominant wavevector components/the incident angle. The DAS output for a given signal~$\vec{s}$ is
\begin{equation}
    y(\phi)=\vec{w}^*(\phi)(\vec{s}+\vec{n}).
\end{equation}
where~$y(\phi)$ is the beamformer output for trial angle~$\phi$,~$^*$ denotes complex conjugate,~the signal $\vec{s}$ is a column vector associated with an unknown angle,~$\vec{n}$ is the additive noise. The weight array is
\begin{equation}
    \vec{w}(\phi)=\begin{pmatrix}
    1 & \exp(\im k_{in} d\sin\phi) &  \exp(2\im k_{in} d\sin\phi)  & \cdots  \exp(\im (n_pn_c-1)k_{in} d\sin\phi) 
    \end{pmatrix}
\end{equation}
where~$k_{in}$ is the incident wavevector,~$d$ is the sensor distance, and~$n_p n_c$ is the total number of sensors. The trial angle~$\phi$ that maximizes~$y$ is the evaluated angle of incidence.

\clearpage

\bibliographystyle{aipnum4-1}
\bibliography{Manuscript}% Produces the bibliography via BibTeX.
\end{document}